\documentclass[conference]{IEEEtran}

\usepackage{algorithm}
\usepackage[noend]{algorithmic}
\usepackage[normalem]{ulem}
\usepackage{multirow}
\usepackage{hhline}
\usepackage{rotating}
\usepackage{diagbox}
\usepackage{subcaption}
\usepackage{xspace}
\usepackage{color, colortbl}
\usepackage{titlesec}
\usepackage{microtype}
\usepackage{booktabs}
\usepackage{tabularx}

\titlespacing\section{0pt}{6pt plus 2pt minus 4pt}{4pt plus 0pt minus 2pt}
\titlespacing\subsection{0pt}{6pt plus 2pt minus 4pt}{4pt plus 0pt minus 2pt}
\titlespacing\subsubsection{0pt}{6pt plus 2pt minus 4pt}{4pt plus 0pt minus 2pt}

\newcommand{\ourSystem}{FaasCamp\xspace}
\newcommand{\newPool}{Reclaim Pool\xspace}

\makeatletter
\newcommand{\linebreakand}{%
  \end{@IEEEauthorhalign}
  \hfill\mbox{}\par
  \mbox{}\hfill\begin{@IEEEauthorhalign}
}
\makeatother

\begin{document}

\title{Caching Aided Multi-Tenant Serverless Computing}

\author{\IEEEauthorblockN{Chu Qiao}
\IEEEauthorblockA{University of Delaware\\
qiaochu@udel.edu}
\and
\IEEEauthorblockN{Cong Wang}
\IEEEauthorblockA{Zhejiang University\\
cwang85@zju.edu.cn}
\and
\IEEEauthorblockN{Zhenkai Zhang}
\IEEEauthorblockA{Clemson University\\
zhenkai@clemson.edu}
\linebreakand
\IEEEauthorblockN{Yuede Ji}
\IEEEauthorblockA{University of Texas at Arlington\\
yuede.ji@uta.edu}
\and
\IEEEauthorblockN{Xing Gao}
\IEEEauthorblockA{University of Delaware\\
xgao@udel.edu}
}


\maketitle


\begin{abstract}
One key to enabling high-performance serverless computing is to mitigate cold-starts.
Current solutions utilize a warm pool to keep function alive: a warm-start can be analogous to a CPU cache-hit.
However, modern cache has multiple hierarchies and the last-level cache is shared among cores, whereas the warm pool is limited to a single tenant for security concerns.
Also,  the warm pool keep-alive policy can be further optimized using cache replacement algorithms.
In this paper, we borrow practical optimizations from caching, and design \ourSystem, a caching-aided multi-tenant serverless computing framework. 
\ourSystem extends the single-tier warm pool into multi-tiers, with a reclaim pool introduced enabling secure function instance sharing among tenants.
Also, \ourSystem leverages machine learning to approximate the optimal cache replacement policy to improve the warm rate.
We have implemented a prototype and conducted extensive experiments under multiple scenarios. 
The results show that \ourSystem can outperform existing platforms with minimal overhead.

\end{abstract}

\maketitle

\section{Introduction}
\label{sec: introduction}
Serverless computing, also known as Functions as a Service (FaaS), is an emerging cloud computing paradigm enabling cost-efficiency and elasticity for high-productive software development.
Serverless computing decomposes computation into functions (e.g. small applications dedicated to specific tasks).
It provides users with great scalability and flexibility by completely hiding the underlying server management, and thus tenants can focus on function development. 
It also allows service providers to efficiently manage hardware resources.
To date, serverless computing has been adopted not only by major cloud platforms~\cite{lambda,google,azure}, but also in resource-limited environments (e.g., edge/mobile servers~\cite{xie2021serverless,nastic2017serverless,baresi2019towards,aslanpour2021serverless}).

One challenge for high-performance serverless computing is to mitigate the effect of cold start, which refers to the process of initiating a new execution instance.
Reducing cold start is essential for both tenants (e.g., user experiences) and vendors (e.g., resource management).
While extensive research efforts (e.g., reducing startup latency~\cite{Catalyzer,sock,shillaker2020faasm} or using a lightweight sandbox~\cite{Firecracker,ao2022faasnap}) have been devoted to optimizing the cold start process, existing cold start start-up latency still ranges from hundreds of milliseconds to several seconds~\cite{faasnet}, whereas a typical serverless function only lasts 1 second~\cite{wild}.
To battle against cold start, one common approach deployed by major platforms is to maintain a warm pool~\cite{faascache,lambwarm,silva2020prebaking}. 
After a function is executed, the underlying sandbox (e.g., container) remains idle in the warm pool and can be reused for the same type of request later (i.e., a warm start).
A warm start avoids building a sandbox from scratch, and thus can save a substantial amount of time.
To further enhance warm pool efficiency, state-of-the-art approaches have treated warm pool as CPU caching: a warm function execution is equivalent to a cache hit, while a cold start is similar to a cache miss.
Based on this insight, previous works such as FaasCache~\cite{faascache} have employed cache replacement algorithms to optimize the warm pool keep-alive policy. 

While modern CPU caching utilizes a tiered memory system (e.g., L1/L2 and last-level cache), existing serverless computing approaches only maintain a single-tier warm pool for each individual tenant. 
Even for the same type of request, a sandbox (e.g., container) used by one tenant in the warm pool cannot be shared by another tenant.
The reason is mainly for security concerns: even though most containers in serverless computing are stateless,
intermediate data can still be stored inside.
However, such a design may not perform well under multi-tenant serverless computing scenarios, especially in a resource limited environment (e.g., edge computing), where the server resources (e.g., memory) are inadequate. 
As the total size of the warm pool is limited, different tenants need to compete for spaces in the warm pool, significantly increasing the number of container evictions. 
A new user will almost certainly get a cold start even if the warm pool contains available containers. 
Unfortunately, most existing approaches are not optimized for resource limited environments.
Meanwhile, existing cache replacement algorithms, including Least Recently Used (LRU), Least Frequently Used (LFU), and Greedy-Dual-Size-Frequency~\cite{faascache}, still cannot achieve optimized performance for serverless computing. 

In this paper, we present \ourSystem, a cache-aided framework for enabling high performance multi-tenant serverless computing. 
The main insight of \ourSystem is to borrow practical optimizations from the modern caching system. 
First, motivated by the hierarchy caching architecture, \ourSystem extends the existing single-tier warm pool to multiple layers, by adding a second tier pool (i.e., reclaim pool).
Different from containers in the warm pool which are specific to tenants, containers in the reclaim pool can be shared for the same type of request among tenants.
Specifically, the reclaim pool leverages the checkpoint and restore technique to enable secured container sharing: every time a new type of function is executed, \ourSystem makes a checkpoint at the beginning, and restores the container when moving it to the reclaim pool so that all intermediate data is erased. 

Additionally, motivated by learning-based cache replacement optimizations~\cite{hawkeye, glider, rlr}, \ourSystem applies deep learning techniques to evict containers guided by the optimal Bélády’s algorithm for CPU caching.
We first develop a simulator to efficiently generate training traces, perform feature engineering, and train a machine learning model.
A neural network based classifier is chosen to approximate Bélády's algorithm.
Then, \ourSystem collects multiple system runtime statistics and per-container information for each warm container.
When multiple warm containers exist in the reclaim pool, \ourSystem feeds the collected information to a loaded deep learning model for inference, and selects the optimal one for eviction. 

We have developed a prototype of \ourSystem on top of OpenWhisk.
Specifically, \ourSystem employs a modular design with several key components incorporated. 
The reclaim pool extends the finite state machine of OpenWhisk so that containers are carefully provisioned among different pools.
\ourSystem further develops a machine learning (ML) based invoker for container dispatching: a StateTracker component collects all desired features and transmits the corresponding state vectors to an independent ML module for inference. 

We conduct extensive real-world trace-driven experiments on a small cluster simulating a resource limited environment, to evaluate the effectiveness of \ourSystem under different scenarios.
The experimental results show that \ourSystem can outperform existing approaches (e.g., vanilla OpenWhisk and FaasCache).
The reclaim pool can increase the warm ratio for multi-tenant scenarios, while the ML-guided invoker can make better container eviction decisions compared with multiple policies (i.e., LRU, LFU, and greedy-dual-size-frequency).
Particularly, \ourSystem can significantly enhance the performance for mobile users, who utilize a provided service a few times and then leave.
We also evaluate the overhead of \ourSystem by using a series of microbenchmarks and measuring overall system utilization. 
The results indicate that \ourSystem incurs minimal overhead.


\section{Background}
\label{sec: background}
\subsection{Serverless Computing}

\noindent\textbf{Serverless Computing Platforms (SCPs)} have largely taken the responsibilities of configuring and managing underlying system resources, compared with traditional cloud computing.
Users only need to specify minimal configurations (e.g., desired memory size and programming language) and upload source code (i.e., function) to the platform.
SCPs then deploy isolated and lightweight sandboxes to serve user requests and run each function.
After the user code is executed, the sandbox is reclaimed/destroyed to free the resources.
A function is usually short-lived: Microsoft Azure observes that 50\% of functions execute for less than 1 second on average~\cite{wild}.
Thus, SCPs typically choose containers (e.g., Docker~\cite{docker}) or lightweight virtual machine (e.g., Firecracker~\cite{Firecracker}), so that they can spawn a large number of sandboxes simultaneously. 
In this paper, we consider Linux containers as the underlying ephemeral execution sandbox, which can achieve near-native performance~\cite{seo2014performance,felter2015updated,morabito2015hypervisors,eder2016hypervisor}.

\noindent\textbf{Cold Start.} Executing serverless functions with low latency is critical for user experience~\cite{boucher2018putting,hellerstein2018serverless,klimovic2018pocket,liu2019e3}.
Thus, container latency especially startup latency is one key metric to enhance performance.
Cold start latency is the time to initialize a new container in response to the request for a function. 
Particularly, when a request arrives, SCP spawns a sandbox instance from a pre-built image with the desired language runtime, which further downloads the required packages and libraries, with user code injected.
While extensive research has been devoted to reducing the cold startup time, the initialization process still takes non-negligible overhead, and can last in the order of seconds~\cite{Catalyzer,sock,shillaker2020faasm}. 

\noindent\textbf{Warm Pool.} 
One common approach for battling the cold start is to maintain a pool of warm containers~\cite{faascache,lambwarm,silva2020prebaking,faasnet}. 
After a function is executed, the underlying container is paused in the warm pool for a period of time without being destroyed immediately. 
As a result, the warm container can be reused for the same type of request later.
The duration for a container to stay in the warm pool is referred to as the keep-alive time.
If no request (of the same type) arrives during the keep-alive period, the warm container is destroyed and the corresponding resources (e.g., memory) are released.
Most existing SCPs adopt a fixed-time keep-alive policy: for example, AWS Lambda used to keep containers alive for 5 minutes~\cite{aws5minute}.
Rather than having a fixed-time keep-alive policy, Shahrad et al.~\cite{wild} proposed a hybrid policy to dynamically set the duration based on the histograms of intervals of consecutive function invocations.

\noindent\textbf{Eviction Policy.}
The warm pool size is limited by the amount of allocated memory.  
If the warm pool is full and a container arrives, either the arriving container or an idle container from the warm pool needs to be evicted (i.e., destroyed).
The container eviction can be analogous to CPU caching: a warm function execution is equivalent to a cache hit, while a cold start is similar to a cache miss. 
Most SCPs adopt the least recently used (LRU) or the least frequently used (LFU) policies as the warm container eviction policy. 
However, both are not versatile enough to handle different workload invocation patterns.
FaasCache~\cite{faascache} then applies the greedy-dual-size-frequency caching policy~\cite{greedy-dual} to evict containers by considering multiple factors like object sizes and frequency.

\noindent
\textbf{OpenWhisk Architecture}.
Many open-source SCP frameworks (e.g., Apache OpenWhisk~\cite{OpenWhisk}, OpenFaaS~\cite{OpenFaaS}, and OpenLambda~\cite{OpenLambda}) have been recently developed to facilitate serverless computing adoption on private clouds and dedicated clusters.
Among them, OpenWhisk has been widely used in academia~\cite{faascache, 9749611, yu2021harvesting, zhang2021faster, 10.1145/3542929.3563468,mvondo2021ofc} and industry (e.g., IBM Cloud Function~\cite{ibm}).
OpenWhisk processes user requests with a logically centralized \textit{controller}, which manages all entities.
The \textit{invoker} is in charge of executing the actions: it actively monitors the message pool in Kafka, and assigns workload to existing containers or spawns new containers accordingly.
By default, OpenWhisk sets the keep-alive time to 10 minutes with LRU as the eviction policy.

\subsection{CPU Caching}
Modern computer systems utilize a tiered memory system to enhance their performance (e.g., increase cache hit ratio).
The lower level cache (e.g., L1 and L2) is small with low access latency, but not shared among cores. 
Instead, the high level cache (e.g., L3) is typically slower, but shared by all cores in a CPU socket. 
Besides the multi-level caching architecture, the cache replacement algorithm is crucial to improve the hit ratio.
While LRU and LFU are common cache replacement policies, extensive research has been conducted to optimize the cache replacement algorithm, from naive heuristic-based solutions~\cite{jaleel2010high,duong2012improving,faldu2017leeway} to sophisticated learning-based solutions~\cite{hawkeye, glider, rlr, wu2011ship, jimenez2017multiperspective}. 

\begin{figure}[t]
\includegraphics[scale=0.85]{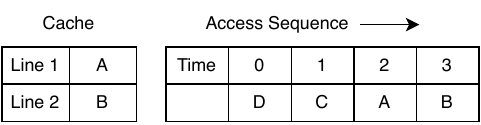}
\centering
 \vspace{-1mm}
\caption{{Bélády's cache replacement policy example: Object B with a reuse distance 4 should be evicted.}}
\label{fig: Belady}
\vspace{-2mm}
\end{figure}

The idea of many learning-based approaches is to train a machine (deep) learning model to learn the optimal policy, which is referred to as Bélády's algorithm~\cite{belady1966study}.
Basically, Bélády's algorithm evicts objects based on the reuse distance, which is the time difference between the current and the next time when the same object is accessed again.
The one with the furthest reuse distance will be evicted first.
Figure~\ref{fig: Belady} illustrates a simple example.
Supposing the cache size is 2, the cache is fully occupied by objects A and B at time point 0. 
Thus, when object D arrives, the cache replacement policy needs to evict one object (e.g., A or B) in the cache. 
According to the access sequence, the reuse distances of A and B are 3 and 4, respectively, as shown in Figure~\ref{fig: Belady}.
As a result, object B is evicted based on Bélády's algorithm.

Bélády's algorithm is infeasible in practice because it requires information from the future.
Thus, recent works focus on developing cache replacement policies to approximate Bélády's algorithm using machine learning methods~\cite{rlr}.
For example, Hawkeye~\cite{hawkeye} and Glider~\cite{glider} build classifiers to predict whether a cache is cache-friendly or cache-averse, where cache-averse lines are evicted with high priority.

\subsection{Multi-Tenant Edge Computing}
Edge computing has been used to offload workloads and reduce latency from a centralized computation approach in modern and next-generation networks (e.g., 5G) and IoT infrastructures.
Unlike the cloud clusters, edge servers typically contain limited resources, but still need to provide multi-tenancy with proper isolation. 
One example is Vehicular Edge Computing (VEC)~\cite{du2018computation,qiao2018collaborative,liu2018computation,tareq2018ultra,chen2015efficient,zhao2017tasks}, where cars need to offload computing tasks to edge clusters for accelerated computing.
With thousands of different automobile manufacturers, even though the offloaded tasks might be similar (e.g., functions with similar/identical packages or developed with same programming languages), they cannot be shared among tenants.
Thus, as resource management is crucial in edge computing, both industry~\cite{edge1,edge2} and academia~\cite{le2020auction,zanzi2018m,wang2020dyverse,willis2014paradrop,csenel2021edgenet} have conducted extensive researches to support multi-tenancy on edge servers.


\section{Motivation}
\label{sec: motivation}
\begin{figure*}[t]
\centering
\begin{minipage}{.32\textwidth}
    \includegraphics[scale=0.45]{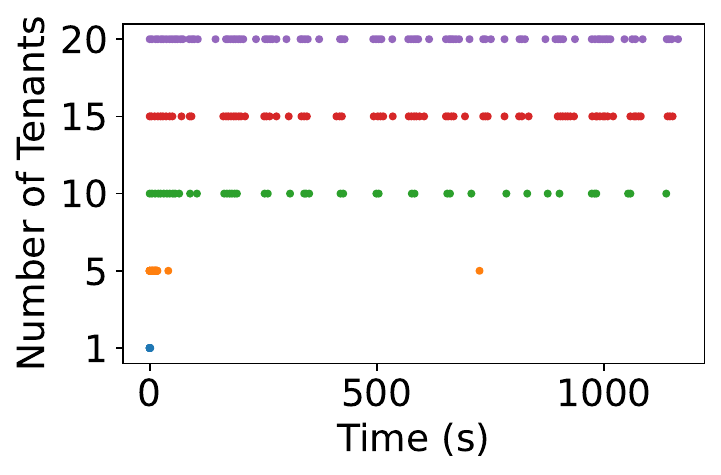}
    \centering
      \vspace{-2mm}
    \caption{Number of cold starts (dots) with respect to tenants.} 
    \label{fig: cold start rate}
\end{minipage}
\hfill 
\begin{minipage}{.32\textwidth}
     \vspace{-3mm}
    \centering
    \includegraphics[scale=0.45]{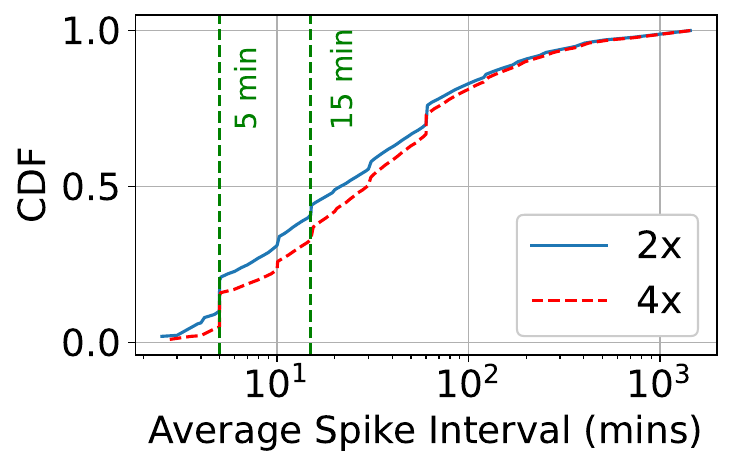}
     \vspace{-2mm}
    \caption{Average Spike Interval.}
    \label{fig: spike}
\end{minipage}
\hfill 
\begin{minipage}{.32\textwidth}
     \vspace{-5mm}
    \includegraphics[scale=0.48]{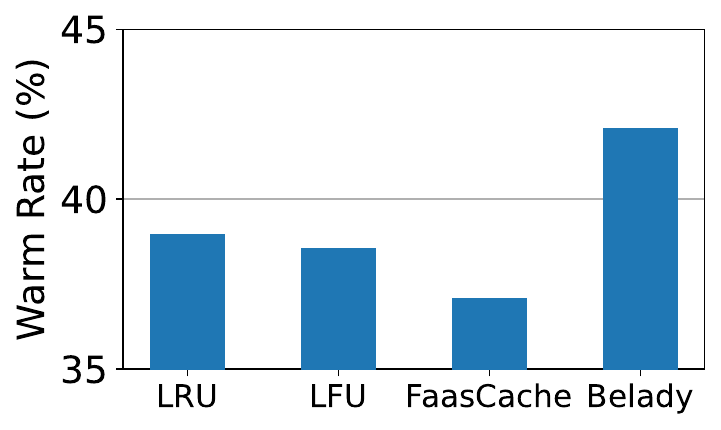}
    \centering
    \vspace{-4mm}
    \caption{Performance Gap.}
    \label{fig: performance gap}
\end{minipage}
 \vspace{-3mm}
\end{figure*}

Existing approaches for maintaining a warm pool of containers works well in cloud computing environments with abundant computing resources.
However, it is challenging to directly adopt it in multi-tenant resource-limit environments (e.g., edge computing). 
There are multiple reasons. First, for security and isolation considerations, a warm pool is separately maintained for each tenant~\cite{peeking}, and warm containers cannot be shared among tenants.
As warm container eviction can be analogous to CPU caching, this design is similar to low-level cache (e.g. L1/L2) which is not shared among cores.

The problem is that, in the edge environment with a high density of users (for example, 5G enables up to thousands of devices existing in one square kilometer), it is impossible to maintain this number of warm pools in edge servers with limited resources.
Thus, in resource limited environments, container eviction happens frequently to make room for new requests.
The performance will be significantly degraded with multiple tenants sharing the same server, as the container of one tenant can evict containers from other tenants.
Second, a good eviction policy is important to keep the warm rate high.
But there is still a non-negligible performance gap between the optimal Bélády's method and existing approaches (e.g., LRU, LFU, and Greedy-Dual~\cite{faascache}).

\noindent\textbf{Preliminary Study Setup.} \label{trace setup}
We have conducted several trace-driven preliminary studies using OpenWhisk to demonstrate the above challenges when deploying a single-tier warm pool design in a multi-tenant scenario.
We use the dataset from Azure Functions~\cite{wild}, which includes a real-world trace of serverless functions running in the Azure cloud for 14 days.
In this trace, each user could deploy multiple apps, and each app may be composed of many functions.
The Azure trace includes the invocation count for each function during each minute of a day (i.e., 1,440 minutes in total), without the detailed timestamp of function invocation.
We thus assume all invocations are evenly distributed in the one minute period. 
A small time adjustment is added to guarantee there are no two requests arriving at the same time. 

Following previous work~\cite{faascache}, we select eight workloads from FunctionBench~\cite{funcbench}.
We create OpenWhisk \textit{actions} (i.e., functions) for each workload with 256 MB memory and pre-built docker image that includes all required packages.
Note that the workload selection is not crucial for this preliminary study, and we expect any functions (e.g., different programming runtimes) will produce similar results.
A detailed introduction of selected workloads are presented in Section~\ref{sec: modelDesign}.

\noindent\textbf{Multi-Tenant Performance Degradation.}
All tenants share the same warm pool and need to fight for memory spaces.
With more tenants using the same server, the warm rate will be decreased due to the competition.
We run an experiment using a fixed trace sequence (e.g., serialize multiple traces as mentioned above) with the same workload, but with  different numbers of tenants.
OpenWhisk by default sets a user-level namespace to each tenant, and containers are shareable within a namespace.
By theory, since all containers are running the same workload, with only one tenant, the majority of cold starts should happen in the beginning when there are no or only a few warm containers.
All following requests should be warm starts.
However, with multiple tenants, warm containers from one tenant may be evicted by other tenants, and thus cold start happens.

In particular, we evenly distribute 100 traces of the same workload to different numbers of tenants (i.e., 1, 5, 10, 15, 20).
Each trace is sampled from the Azure dataset with 3,034 invocations in total.
The experiment is conducted on our local Linux server (running Ubuntu 20.04) with 16 GB RAM and Intel i5-6600K 3.50 GHz CPU.
OpenWhisk is allocated 2 GB of memory, and the workload is configured with a maximum of 256 MB.
Thus, at most eight containers can exist at the same time.

Figure~\ref{fig: cold start rate} shows the number of cold start with respect to time.
Each colored dot in the figure represents a cold start and each line represents one scenario (e.g., different numbers of tenants).
For a better visualization, we omit some warm starts in sparse regions to shorten the time axis.
In all scenarios, many cold starts happen in the beginning, indicating that the system keeps provisioning containers.
After the initial spike, the number of cold starts decreases as there are available containers in the warm pool.
Overall, with more tenants, more cold starts happen as the system needs to provision new containers due to container evictions.
The situation will become worse with more types of workload involved. 

\noindent\textbf{Invocation Spikes.}
Invocation spikes will further reduce the warm rate in a multi-tenant resource-limited scenario.
In serverless computing, user requests may not arrive evenly. Instead, the bursty invocation pattern (e.g., flash events) is quite common in the wild~\cite{faasnet}.
It is entirely possible that one tenant receives many requests within a short period of time, and becomes idle then.
In this case, the warm pool will be swarmed by containers of one tenant (caused by invocation spikes), and containers of other tenants will be evicted. 
As a result, all the following requests require a cold start. 

We analyze the spike pattern using the Azure dataset~\cite{wild}.
We define a spike as the time point when the number of invocations is at least twice as the average counts.
Figure~\ref{fig: spike} shows the CDF of the average spike interval in minute.
The result shows that 80\% of the spikes have the interval larger than 5 minutes and at least 50\% of the interval is larger than 15 minutes.
It indicates that if the warm pool is not optimized for handling flashing events, many cold starts can potentially happen due to invocation spikes.

\noindent\textbf{Performance Gap.}\label{gap}
With the cache analogy, algorithms from caching can be customized to enhance warm rate in serverless computing.
The optimal solution, Bélády's algorithm, still outperforms existing methods (e.g., LRU, LFU, Greedy-Dual from FaasCache~\cite{faascache}) with a non-negligible performance gap. 
We conduct an experiment to demonstrate this gap.
we use OpenWhisk on commit 3e5b8b2, which utilizes LRU as the default policy. 
We implement a LFU version on top of OpenWhisk.
We also run FaasCache (available through GitHub).
The warm rate of Bélády's algorithm can be calculated with a serialized trace sequence.
The performance of Bélády's algorithm depends on how far it can look into the future: looking ahead to a large number of requests may actually diminish benefit returns~\cite{hawkeye}.
Thus, we empirically choose 30 as the window size to look ahead.
When calculating, we consider the running time of containers: a busy container will not be considered for eviction. 
We also drop incoming requests if all containers are busy and treat them as cold starts. 
Using the exact same trace on different algorithms, Figure~\ref{fig: performance gap} clearly indicates that there is still a performance gap between existing approaches and Bélády's algorithm.
We should mention that the real gap should be even larger, as many requests are dropped by us and considered as cold starts.


\section{System Overview}
\label{sec: overview}
This section presents the overview of \ourSystem.
The main insight of \ourSystem is to borrow practical optimizations from the caching system. 
Particularly, similar to the tiered caching architecture used by modern computer systems, \ourSystem extends existing single-tier warm pool to multiple layers, by adding a second tier pool (i.e., reclaim pool) enabling secure container sharing among tenants.  
Additionally, motivated by learning-based cache replacement optimizations, \ourSystem applies deep learning techniques to evict container guided by the optimal Bélády's algorithm. 

\begin{figure}[t]
\centering
\includegraphics[width=0.85\linewidth]{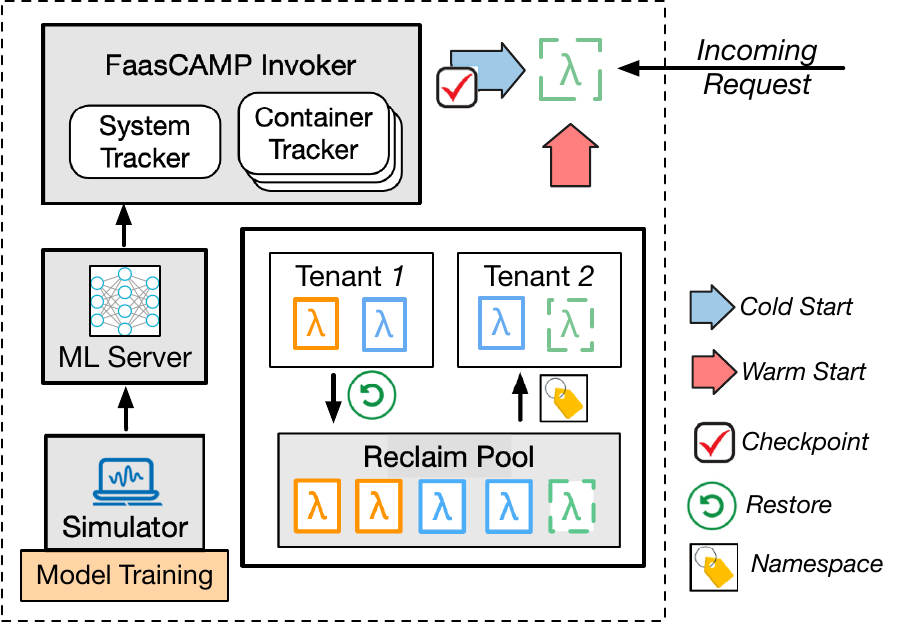}
 \vspace{-1.5mm}
\caption{FaasCamp System Overview.}
 \vspace{-3mm}
\label{fig: overview}
\end{figure}

Figure~\ref{fig: overview} illustrates the high-level architecture of \ourSystem, which adopts a modular design with several key components.
\newPool is the second tier pool that extends the finite state machine (FSM) of vanilla OpenWhisk so that containers in the reclaim pool can be reused by the same type of request from different tenants.  
The \textit{StateTracker} component collects multiple system runtime statistics, and per-container information for each warm container.
Both are fed into a deep-learning based invoker. 
When multiple warm containers (i.e., more than one) exist in \newPool, FaasCamp invoker calculates a score for each container instance, and selects the optimal one for eviction. 
The detailed machine learning model is loaded in a separate container and communicates with the invoker, so that the model can be updated without interfering with the system.
To facilitate the training process, we have developed a simulator to generate training traces and conduct the feature engineering job.
While \ourSystem is built on top of OpenWhisk, the idea can be easily applied to other serverless computing platforms.

\subsection{\newPool Design} \label{design}
\ourSystem extends the existing warm pool into a multi-tier architecture, by adding a reclaim pool where containers can be shared among tenants.
The idea is similar to the last-level cache (LLC), which is shared among all cores.
However, we need to take multiple factors into consideration. 
The reason that the warm pool is limited to a single tenant is for security concerns~\cite{peeking}. 
Even though most containers in serverless computing are stateless, intermediate data can still be stored inside containers~\cite{refunction}. 
Thus, to enable sharing containers among tenants, it is critical to ensure that the container has been safely cleaned: a previously executed function cannot affect future functions, and the currently running function should not infer previous functions.

The overall idea of the reclaim pool is to extend the FSM of Openwhisk by employing the checkpoint-restore technique for ensuring security. 
Every time a new type of function (which needs a new container image) is executed, \ourSystem makes a checkpoint at the beginning. 
This checkpoint ensures that the container is stateless at this point regardless of whether the function is stateful or stateless.
Specifically, the checkpoint procedure will run in a separate process to reduce system overhead and function waiting time. 
Plus, \ourSystem only needs to create checkpoints for new (e.g., checkpoint does not exist) and unique workloads (e.g., requires specific extra libraries in containers). Thus, the total number of checkpoints stored in a server is limited.

After executing the function, the container will be moved to the warm pool as usual. 
If the container has been idle for a period of time in the warm pool (e.g., 5 minutes), \ourSystem moves this container into the reclaim pool by two operations. 
It first removes the ownership (e.g., tenant information) of this container.
Then the container will be restored to the previous checkpoint so that all intermediate data is erased. 
For a new incoming request, \ourSystem will first check the warm pool, and then the shared reclaim pool for warm instances. 
If a warm container exists in the shared reclaim pool, \ourSystem adds user-specific ownership to the container.

\ourSystem does not reserve any memory space for the reclaim pool. Instead, administrators can set the upper limit (e.g., memory size) for the number of containers in the reclaim pool.
For example, the size of the reclaim pool can change dynamically from 0 to 25\% of the total memory allocated to OpenWhisk.
If the system supports at most 16 containers, at most 4 containers can be placed in the reclaim pool.
The specific implementation details (e.g., extending container FSM in Openwhisk) are presented in Section~\ref{sec: implementation}.

\begin{table}
        \captionof{table}{State Vector Features.}
            \vspace{-2mm}
        \label{table: SV}
        \footnotesize
        \hspace{-2mm}
        \begin{tabular}{@{}c@{}|@{}l|@{}l@{}}
        \toprule
        \textbf{Category~} & 
            \textbf{~Feature} & 
            \textbf{~Description} 
            \\ \midrule
        \multirow{3}{*}{System} 
            & ~Current request & ~The workload ID of the current request \\ 
            & ~PII 1, 2 & ~The intervals between past 2 requests \\  
            & ~Past 200 & ~The history of past 200 requests \\ \midrule
        \multirow{7}{*}{Container~}  
            & ~Workload ID & ~The ID of the workload \\ 
            & ~Idle time rank & ~The ranking of idle time among all containers \\ 
            & ~Frequency & ~Total number of container being accessed \\ 
            & ~Frequency rank & ~The ranking of freq. among all containers \\ 
            & ~Alive count & ~\# of requests passed since the first invocation \\ 
            & ~Warm count & ~\# of requests passed since the prev. invocation \\ 
            & ~Past \{10, 50, 100\} & ~\# of invocation of past 10, 50, 100 requests \\ \bottomrule
        \end{tabular}
         \vspace{-0mm}
\end{table}

\subsection{Machine Learning Based Invoker} 
While machine learning has been used to optimize the cache replacement policy, the models were never deployed inside real CPUs due to the obviously high overhead and resource restriction.
Instead, with a similar design principle as cache, the container eviction policy serves as an ideal place for implementing learning-based optimizations, as serving serverless requests are much more resilient (and slow) than object caching in hardware. 

While the idea is simple: developing a machine (deep) learning model to approximate the optimal Bélády's algorithm, \ourSystem still needs to address several challenges. 
First, the performance of machine learning models heavily depends on the quality and size of training data, as well as the selection of features.
The feature engineering work is tedious, requiring many repetitive attempts to try different combinations. 
It is impractical to run workloads in a real server for months just to collect training data and extract corresponding features for model training.
Thus, \ourSystem first adopts a similar approach as FaasCache by implementing a trace-driven discrete event simulator for feature engineering and training data generation.
Based on the simulation, \ourSystem implements \textit{StateTracker} to collect various system information (e.g., request history) and a wide-spectrum of container-dependent states (e.g., invocation frequency and idle counts).
\textit{StateTracker} constructs state vectors for each container, and transmits them to a machine learning server. 
A TensorFlow model performs inferences on all state vectors, and decides the container to evict.
The machine learning model runs in a separate container, and delivers the decision to the invoker, which is responsible for finally evicting the target container. 
Such a design enables \ourSystem to update the machine learning model online without re-compiling the invoker. 
For example, combining the simulator, \ourSystem can easily re-train the model based on recent request patterns (e.g., request distributions).

\section{Machine Learning Model Design}
\label{sec: modelDesign}
This section presents the detailed design of our machine learning model. 
We first implement a trace-driven simulator to emulate the warm pool activities. 
The simulator uses two lists to represent a busy pool of containers and a free (warm) pool.
The simulator supports different numbers of containers in the warm pool, and various workloads with configured running time and cold start latency. 
Given an incoming request from traces, the simulator calculates the end running time based on arrival time, execution duration, and cold start time.
Multiple eviction policies (e.g. LRU, LFU, FaasCache, our machine learning model) can run in our simulator, which further decides whether to evict an idle container in the warm pool or discard the request.
With a fixed sequence of traces, we can also obtain the optimal choice from Bélády's algorithm.
The simulator is written in Python with around 700 LoC, and significantly shortens the time to run the whole sequence in a real server. 
Note that the simulator is not intended to precisely replicate OpenWhisk.
Instead, the simulator enables us to quickly generate training samples for selecting features and comparing different ML algorithms and models.
Therefore, algorithms (e.g., FaasCache, LRU, and LFU) might perform differently between the simulation and the actual system.

\noindent\textbf{Setup}. We select eight workloads from FunctionBench~\cite{funcbench}, including basic computation operations (e.g., float operation), complex mathematics calculations (e.g., linear equation solving, matrix multiplication, and AES encryption), machine learning workloads (e.g., model training, model serving), and real-world applications (e.g., HTML rendering, image processing). 
We use the same method described in Section~\ref{trace setup} to generate traces. 
As the Azure dataset includes a significant amount of data from 14 consecutive days, we randomly select a portion of data (roughly 4,400 traces) from the first day as the training data.
Particularly, some traces contain overmuch invocations (e.g., thousands per minute) causing the training process inefficient.
Meanwhile, if we select multiple traces with very few invocations (e.g., less than 10 per day), the generated data might not be enough for model training.
Thus, we mark these traces as outliers, and conduct experiments without them (i.e., selecting traces from 10 to 10k).

\begin{figure*}[t]
\centering
    \includegraphics[scale=0.35]{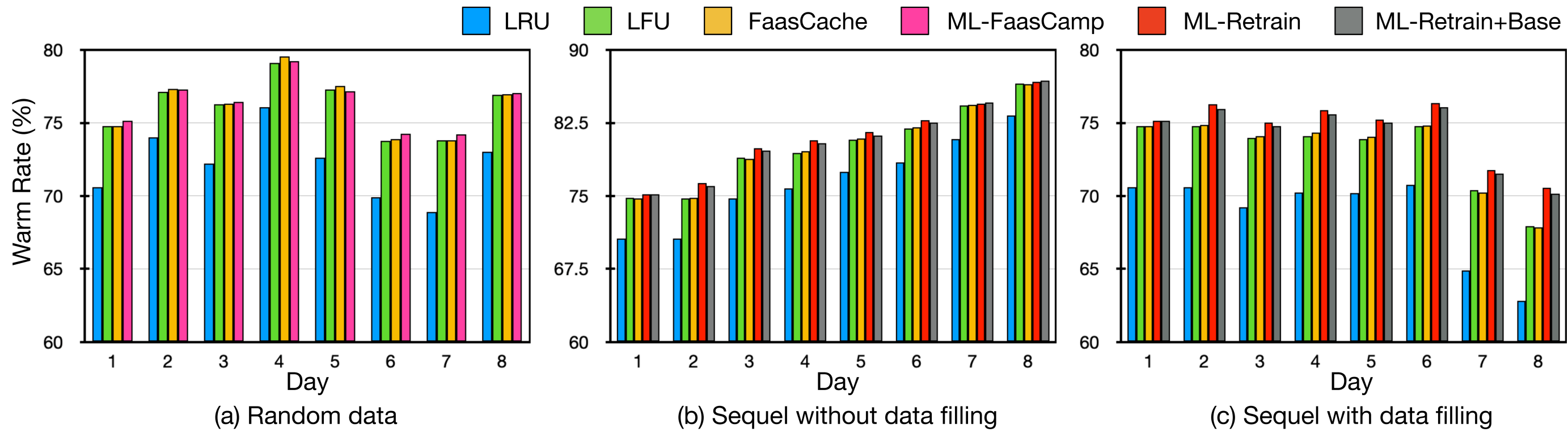}
 \vspace{-1.5mm}
\caption{Simulation Results.}
 \vspace{-3.5mm}
\label{fig:model}
\end{figure*}

Our plan is to deploy the ML based invoker on the reclaim pool, which is simply a portion of the entire warm pool. 
Thus the size is limited in a resource-limited server. 
Each time, we randomly feed 40 traces into our simulator, with state vectors (e.g., features) and labels (based on Bélády's algorithm with 30 requests into the future) generated.
We only use day 1's data for training.
State vectors and labels are input-output pairs to train the machine learning model.
We randomly generate 100 sets (each set contains 40 traces) as our training data.
Because the trace length varies in a large range, even with a fixed number of traces (i.e., 40), the input-output training pairs for each set still significantly differ from each other. 
Note that these 100 sets of data are only used for model training, and we generate new data for model evaluation (as described later).

\subsection{Eviction Model Details} \label{ML design}
This subsection presents the details of \ourSystem's ML architecture, with its choice and rationale described. 
Our exploration might not be the optimal choice as it can change in different scenarios. 
Particularly, we use the simulation results to show that these features have the potential to outperform traditional non-learning based methods.
The detailed evaluation is conducted on real servers in Section~\ref{sec: evaluation}.

\subsubsection{Features.}
As Table~\ref{table: SV}, \ourSystem uses two different types of features: one set describing the overall state information of the warm pool (i.e., system state vector), and another set describing the information of particular workload (i.e., container state vector). 

\noindent\textbf{System state vector} provides the historical information for the server (e.g., past 200 requests).  
The \textit{rationale} is that request invocations might exhibit similar patterns in a short amount of time. 
The past information can serve as an estimate of the future invocation pattern, which is learned by Bélády's algorithm.
For each type of workload, we also record past invocation intervals (PIIs), indicating the number of requests between the two requests on the same workload.
As the eviction happens when a new request arrives, the interval is described as the number of requests, rather than a period of time.
Specifically, we track two most recent intervals (e.g., PII\textsubscript{1} and PII\textsubscript{2}) for each type of workload. 
For example, if the current request is for workload A, PII\textsubscript{1} is the number of fulfilled requests \textit{since}  A was served last time, and PII\textsubscript{2} is the number of requests \textit{before} A was served last time.

\noindent
\textbf{Container state vector} contains per-container information required to make a replacement decision.
Multiple classes of features are extracted for each container, such as the current access information (e.g., workload type), the frequency information of each type of container, and the amount of idle time stayed in the warm pool. 
We also collect the ranking information (e.g., frequency rank) as features.
In addition to the overall historical information in the system state vector, we also track the number of historical invocation for each type of container.
Statistics are updated on every request: the container's alive count is reset to zero if it is newly created, and is incremented on every incoming request to the system. 

\subsubsection{ML Architecture.}
When designing the ML architecture, one of the challenges  is to select a model that is general to different configurations.
For example, the model should have competitive performance with different memory sizes of the warm pool.
Also, the number of available containers is dynamic in the warm pool: we cannot evict a busy container while it is serving a request. 
In both cases, the varying number of containers results in different lengths of inputs, which contradicts a multi-class neural network classifier with a fixed-length input.
We have explored multiple choices, including deep neural networks and convolutional neural networks with zero padding, as well as recurrent neural networks for time series classification, but none of them are satisfactory.  

Finally, we perform a prediction for each available container instance to decide whether it is warm-friendly or warm-averse.
\textit{Warm-friendly} indicates that this type of container should be kept in the warm pool with high priority, and \textit{warm-averse} container is with low priority.
Particularly, we train a neural network with 5 hidden layers and 256 nodes in each layer, using the binary cross-entropy loss function. 
For each container, the model utilizes the overall system state vector and the corresponding container state vector, and outputs a probability that this container should be evicted. 
Then, the container with the highest probability is evicted.

\subsection{Model Performance}
We evaluate our model (via the simulator) using newly generated datasets.
We randomly generate another 20 sets of traces (each set contains 40 Azure traces), ensuring that the testing data is unseen by the model.
Particularly, we randomly generate such testing data for eight days (i.e., days 1-8) independently.
This allows us to test the feature generality, as the model is trained using only day 1's data. 
Figure~\ref{fig:model}(a) shows the results of our machine learning model compared with LRU, LFU, and FaasCache.
Our machine learning model outperforms them on day 1's data. 
This is expected as our model is also trained with day 1's data.
Also, our model can achieve competitive (even the best) performance in the following days, indicating the potential and effectiveness of our selected features.

\noindent\textbf{Model Updating.}
One advantage of our design is that the model can be updated frequently based on recent request patterns. 
For example, in a realistic scenario, after deploying the system for one day, we can update our model based on the collected historical data. 
To simulate this scenario, we adopt the following method to generate testing datasets: we keep day 1's newly generated testing traces unchanged, and  further extract data of the same traces from the following days (e.g., days 2-8) for evaluation. 
In some traces, requests might disappear in the next day.
This is quite normal as one tenant might leave the service on the other day.
Thus, we adopt two strategies on these traces: (1) we ignore the missing requests; (2) we randomly add more traces to the evaluation data to make them 40.

We further adopt two model retraining strategies.
First, when evaluating the performance of the next day (e.g., 3rd day), we retrain the model using the data of the previous day (e.g., 2nd day) in the testing trace. 
Second, to avoid the catastrophic forgetting phenomenon (the retrained model performs poorly on the old data because the new data may have a different distribution), we mix the data from the base training traces (i.e., from day 1) and day 2 testing traces to retrain and evaluate the new model with data from day 3. 
The same approach also applies to other days.

In both cases, model retraining can outperform all previous methods (i.e., LRU, LFU, FaasCache). 
With a modular design, \ourSystem allows administrators to update the machine learning model for achieving satisfactory performance. 


\section{Implementation Details} 
\label{sec: implementation}
This section describes some non-trivial implementation details for integrating \ourSystem on top of OpenWhisk.
The major modification is to develop our own \textit{invoker} to dispatch workloads and manage the life-cycles of containers. 

\subsection{Reclaim Pool}
In the vanilla OpenWhisk, the \textit{invoker} manages containers through \textit{ContainerPool} and \textit{ContainProxy}.
\textit{ContainerPool} oversees two pools, \textit{busyPool} (contains busy containers) and \textit{freePool} (i.e., warm pool).
\textit{ContainProxy} is a FSM created for each newly spawned container.
The state changes when the corresponding container moves to another pool. 
\ourSystem extends the FSM transition and the logic of \textit{ContainerPool} to support the second tier pool (\newPool).  

\noindent\textbf{Checkpoint and Restore Optimization.}
We adopt the CRIU~\cite{criu} project for container checkpoint/restore functionality.
CRIU creates checkpoints by gathering information of the container processes from the \texttt{/proc} file system, and restores the processes through the \texttt{fork()} system call with previously collected data.
The data created after the checkpoint will be removed when the container is restored, and thus the container can be safely used by a new user.

\begin{figure}[t]
\includegraphics[width=0.75\linewidth]{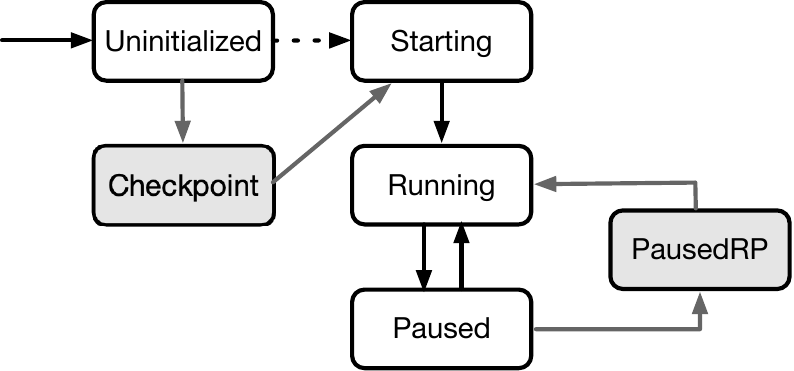}
\centering
\caption{Container Finite State Machine.}
 \vspace{-3mm}
\label{fig:container state}
\end{figure}

When a new container is created, we add another state, \textit{Checkpoint}, in the \textit{ContainProxy} (i.e., FSM).
The detailed FSM is illustrated in Figure~\ref{fig:container state}, where states marked with the grey color are added by FaasCamp.
If a checkpoint is needed (e.g., no previous checkpoint exists for this type of workload), \ourSystem builds a checkpoint by spawning an extra container in another thread, while the original one follows the procedure in the FSM and is provided to the user.
The extra container is not counted towards the container limit in the OpenWhisk container pool, and will be removed after the checkpoint is successfully created.
This method, although consumes extra memory, allows us to cause minimal impact on users.
Also, the number of checkpoints is determined by the number of unique workloads, thus generating them is often a one-time effort.

Containers are restored when they are moved from the free pool to the reclaim pool.
Previously the container enters into a \textit{Paused} state for staying in the warm pool, where all processes of this container are suspended.
We add an extra state \textit{PausedRP} for containers staying at the reclaim pool.
The restore process takes place when the state transitions from \textit{Paused} to \textit{PausedRP}.
This process requires to shutdown the container first, and then restore/restart the container. 
Although the restore process causes a cold start, this cold start is hidden from users (as it is not happened for serving a request).

\noindent\textbf{Namespace Handling.}
OpenWhisk add a user-level namespace to all containers ensuring that warm containers are isolated among tenants.
When moving a container between pools, \ourSystem also modifies the namespace.
Specifically, we add a namespace for the reclaim pool, and label containers with this namespace when they enter into the reclaim pool, indicating the container is free to all tenants.
The namespace is set to the specific tenant if the container is taken to serve new a request (i.e., becomes busy).
Compared with using containers in the warm pool, the namespace operation is the only overhead added for containers in the reclaim pool.
This overhead is negligible as we evaluate it in Section~\ref{sec: evaluation}.

\begin{algorithm}[t]\label{algorithm: container provision}
\caption{Container Provision}
\small
\label{alg:container}
\begin{algorithmic}[1]
\IF{WP container $>$ 0 or RP container $>$ 0}
    \STATE Assign a free container to the new request
\ELSE
    \STATE maxFreeSpace = maxContainerCount - busyPool.size
    \IF{maxFreeSpace == 0}
        \STATE Send the current request to buffer
    \ELSE
        \STATE freeSpace = maxFreeSpace - WP.size - RP.size
        \IF{freeSpace $>$ 0}
            \STATE createContainer()
        \ELSIF{WP.size $>$ 0}
            \IF{RP $>$ 0}
                \STATE Remove a container from RP
                \STATE createContainer()
            \ELSE
                \STATE Remove a container from WP
                \STATE createContainer()
            \ENDIF
        \ELSIF{WP.size == 0}
            \STATE Remove a container from RP
            \STATE createContainer()
        \ENDIF
    \ENDIF
\ENDIF
\end{algorithmic}
\end{algorithm}

\noindent\textbf{Container Provision Algorithm.}
One challenge for implementing the reclaim pool is to ensure that the total number of containers in all pools does not exceed the memory limit in the whole time.
These include busy containers in the busy pool and idle (warmed) containers in both warm and reclaim pools. 
The invoker needs to carefully handle the interaction (e.g., move a container) among different pools when provisioning containers and finishing workloads. 
Algorithm~\ref{alg:container} presents the details of how \ourSystem reacts to a new request.
The \textit{invoker} searches idle containers in the warm pool and reclaim pool (Lines 1-2).
If there is no available container, our invoker needs to create a new container if there is still space.
First, the invoker checks if the memory space is entirely occupied by running containers.
If the number of running containers is already reached the maximum limit, no memory is available for serving the current request.
Thus, this request will be buffered in a queue for storing unfulfilled requests (Lines 5-6).
Next, the invoker checks whether all free spaces are occupied by warm containers (in both warm and reclaim pools).
if not, a new container can be directly created (Lines 9-10).
Otherwise, FaasCamp needs to evict an idle container to make space for serving the current request.
\ourSystem prioritizes reserving space in the warm pool, so the invoker will first check and evict idle containers in the reclaim pool (Lines 12-14 and 18-20).
Instead, if all idle containers are in the warm pool (Lines 15-17), a warm pool container will be destroyed. 
When a container exits the busy pool (e.g., finish serving a request), this container will be placed in the warm pool (only belongs to the corresponding tenant).
Meanwhile, if the warm pool is already full, we will select and move a container from the warm pool to the reclaim pool.

\begin{figure*}[t]
    \begin{minipage}{0.33\textwidth}
        \captionof{table}{Reclaim Pool Performance.}
        \scriptsize
        \begin{tabular}[t]{@{}c|c|c|c@{}}
        \toprule
            \multirow{2}{*}{\textbf{System}} & 
                \multicolumn{3}{c}{\textbf{Avg. Warm Rate (win counts)}}\\\cmidrule(l){2-4} 
                & \textbf{S1} & \textbf{S2} & \textbf{S3}\\\midrule
            OpenWhisk (32-0) & 
                85.43 &
                10.25 &
                81.17 \\
            \ourSystem(24-8) & 
                87.92 (15) &
                36.12 (15) &
                85.28 (15) \\
            \ourSystem(26-6) & 
                86.42 (13) & 
                27.27 (15) & 
                82.98 (13) \\
            \ourSystem(28-4) & 
                85.85 (11) &
                18.36 (15) &
                81.88 (11) \\
                \midrule
            OpenWhisk(24-0) & 
                78.8 &
                9.89 &
                77.41 \\
            \ourSystem(16-8) & 
                82.7 (15) &
                39.92 (15) &
                82.14 (14) \\ 
            \ourSystem(18-6) & 
                80.95 (15) &
                29.91 (15) &
                78.42 (13) \\
            \ourSystem(20-4) & 
                79.65 (11) & 
                18.1 (15) & 
                77.63 (10) \\ 
            \bottomrule
        \end{tabular}
        \label{tbl: rp}
    \end{minipage}
\hspace{0mm}
\hfill
    \begin{minipage}{0.32\textwidth}
        \includegraphics[width=0.92\linewidth]{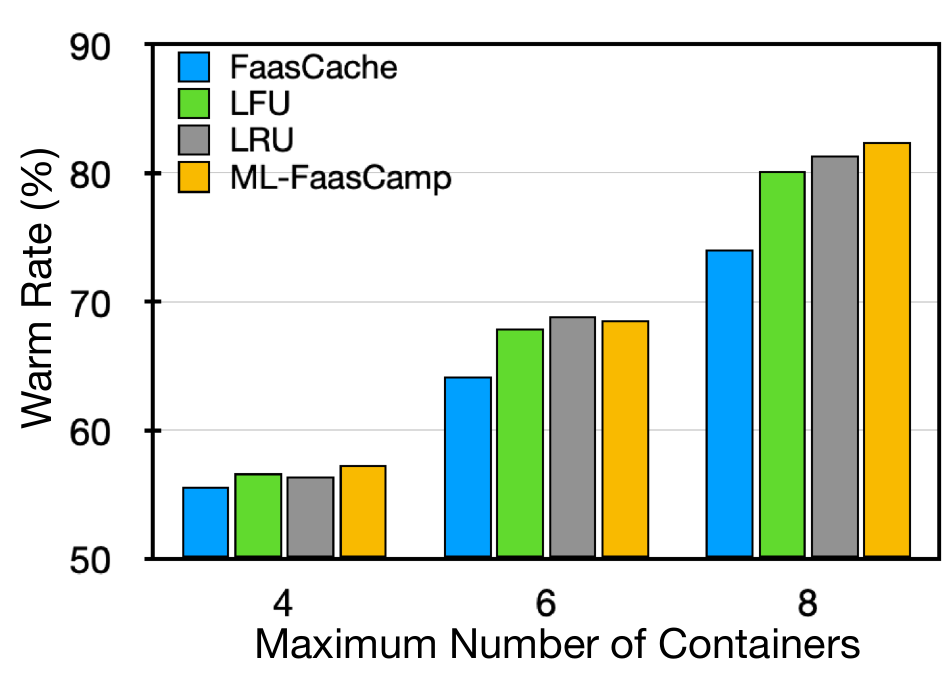}
        \centering
        \caption{Eviction Policy Comparison.}
        \label{fig: exp2}
    \end{minipage}
    \hfill
        \begin{minipage}{0.32\textwidth}
        \includegraphics[width=1\linewidth]{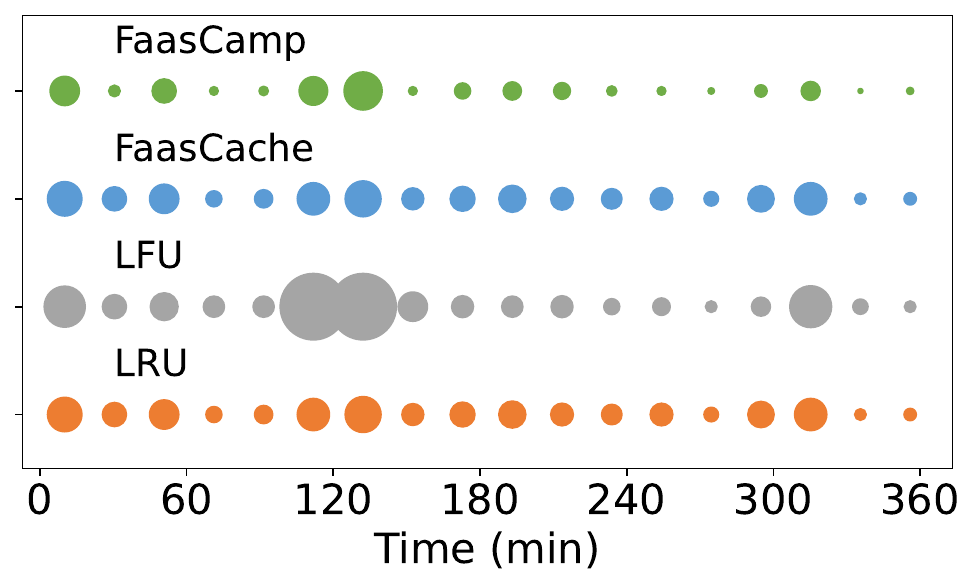}
        \centering
        \caption{\ourSystem Overall Performance (circles indicate cold starts).}
        \label{fig: 6 hrs exp}
    \end{minipage}
     \vspace{-3mm}
\end{figure*}

\subsection{ML Module} \label{modular design}
The high-level idea of our ML module has been introduced in Section~\ref{sec: overview}.
We add a \textit{StateTracker} in the invoker to collect all desired features and transmit the corresponding state vectors to an independent ML server for inference.
The invoker then evicts one container based on the results from the ML server. 
Currently, we only evict containers using machine learning in the reclaim pool, while it can be easily extended into the warm pool as well.

\noindent\textbf{StateTracker.}
The \textit{StateTracker} contains two components, \textit{SystemTracker} and \textit{ContainerTracker}.
\textit{SystemTracker} is initialized with the \textit{invoker} and tracks all invocation history. 
It also manages and triggers \textit{ContainerTrackers} to update statistical data (e.g. increment idle count).
For each newly spawned container, a \textit{ContainerTracker} is created to track all container features.
Each container tracker will be removed when the corresponding container is destroyed.

\textit{SystemTracker} is responsible for generating state vectors when a container needs to be evicted.
It constructs state vectors by gathering information from each \textit{ContainerTracker} and calculating PIIs using the recorded invocation history.

\noindent\textbf{ML Server}.
The ML model is trained using TensorFlow with Python in the simulator.
OpenWhisk is developed with Scala, which is not officially supported by TensorFlow, so we cannot directly integrate the ML model in the invoker.
Thus, we seek a modular design by creating an independent ML server in a separate container. 
While multiple methods (e.g., shared memory) can be used for information exchange among two containers, currently we utilize network sockets to transfer state vectors and model decisions.
The overhead of data transfer is not the bottleneck (evaluated in Section~\ref{sec: evaluation}). 
The model computes a score for each idle container, and returns the one with the highest score for eviction.
In this way, for a cluster with multiple nodes, each node can deploy its own ML server for independent inferences, or rely on a centralized ML server by transmitting state vectors.


\section{Evaluation}
\label{sec: evaluation}
\subsection{Experimental Setup} \label{eval: method}
Real-world serverless computing typically contains a wide-range number of requests from different users, from single digit to hundreds of thousands per day.
For example, in the Azure trace, 47\% of users only have 1 function deployed, and over 60\% of the functions are only invoked once per day.
We thus consider two types of users to emulate real-world scenarios. 
The first type is \textit{regular users} who deploy their applications on the server with frequent invocations.
The second type is \textit{mobile users}. Such users might visit this location, utilize the provided services (e.g., from an edge server) a few times and then leave. 
Based on the characteristics of both users, we divide the dataset into two sections: mobile users are assigned with traces that have less than 100 invocations, and the rest are regular users.

We further consider three scenarios:\vspace{-0.5mm}
\begin{itemize}
    \item \textbf{Regular User Only (scenario 1):} all tenants are regular users who might run multiple different applications. So each tenant has multiple traces assigned. 
    \item \textbf{Mobile User Only (scenario 2):} all tenants are mobile users, and each tenant has only one trace assigned.
    \item \textbf{Mixed (scenario 3):} Both mobile and regular users exist with different ratios (e.g., 1:1 or 3:1).\vspace{-0.5mm}
\end{itemize}

To reduce the impact of trace selection (the performance is not tied to specific traces), we first conduct each experiment with different traces (randomly generated) multiple times but over short periods.
Particularly, for evaluating reclaim pool (\S~\ref{eval: newPool}) and ML-based invoker (\S~\ref{eval: ML}), we randomly select a short period (10$\sim$15 minutes), while keeping a wide range of invocation counts. 
Sampling mobile users might cause 0 invocation in our randomly selected period. 
In this case, one invocation is randomly added.
We also conduct a long-term trace-driven experiment to evaluate the performance (\S~\ref{eval: overall}) and overhead (\S~\ref{eval: overhead}) of FaasCamp on a single server.
Finally, we evaluate FaasCamp in a small cluster (\S~\ref{eval: cluster}).

\subsection{Effectiveness of \newPool} \label{eval: newPool}
We first compare the warm rate of the reclaim pool against vanilla OpenWhisk, with the default LRU as the eviction policy for both systems. 
With each container's memory limited in 256MB, we conduct experiments with two setups: (1) 6G maximum memory with 24 containers and (2) 8G maximum memory with 32 containers in total. 
These setups ensure a reasonable representation of a server with limited resources. 
For each setup, we further test two different configurations by adjusting the memory limit of the reclaim pool.
For example, for the 8G setup, we set idle containers limit in the reclaim pool to 4 or 8. 
The details of our configuration are presented in Table~\ref{tbl: rp}: FaasCamp (24-8) indicates that at most 24 and 8 containers can exist in the warm pool and reclaim pool, respectively.
For each setup, we conduct 15 experiments, with 360 experiments performed in total.

Table~\ref{tbl: rp} shows the results: \ourSystem robustly outperforms vanilla OpenWhisk on almost all experiments. 
Particularly, the reclaim pool can greatly increase the warm start ratio for mobile users (scenarios 2 and 3).
In some cases, the increase can reach 4x better.
The reason is obvious: with less invocations, mobile users will almost certainly get cold start, while the reclaim pool enables them to use idle containers from other tenants.

\subsection{ML Eviction Accuracy} \label{eval: ML}
Next, we compare the warm rate of our ML-based invoker with other eviction algorithms (i.e., LRU, LFU, and FaasCache). 
To reduce the impact of tenants, we distribute all traces to only one tenant, so that containers will not be evicted due to the impact of cross-tenant.
We conduct experiments 30 times for each setup (360 experiments in total) using random traces from the regular users.
As FaasCamp only adopts the ML eviction in the reclaim pool, we evaluate different maximum container counts (i.e., 4, 6, 8), which are the sizes evaluated in \S~\ref{eval: newPool}.
Note that the machine learning model is trained with 8 maximum containers.
Also, the reclaim pool is not enabled.
Figure~\ref{fig: exp2} shows the results, and the bar represents the average warm rate of all experiments.
Our ML model clearly outperforms all others policies in the 8 container case (e.g., 11\% better than FaasCache). 
Meanwhile, even for other cases, the model still exhibits competitive performance (e.g., the best one for the 4 container case). 
The result shows the generality of our model and selected features.
We should emphasize that, system administrators can always train the model based on their setups (e.g., number of maximum containers) for best performance.

\begin{figure}[t]
\begin{subfigure}{.49\linewidth}
    \hspace{-3mm}
    \includegraphics[scale=0.6]{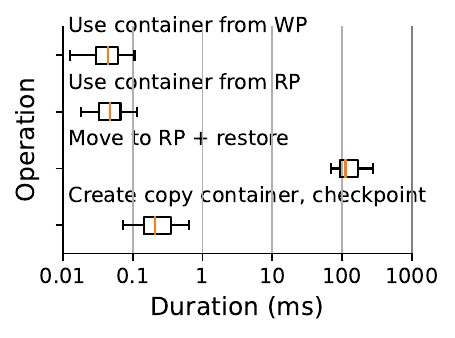}
     \vspace{-6.5mm}
    \caption{Reclaim pool related}
    \label{fig: rp_ops}
\end{subfigure}
\hfill 
\begin{subfigure}{.49\linewidth}
    \centering
    \includegraphics[scale=0.6]{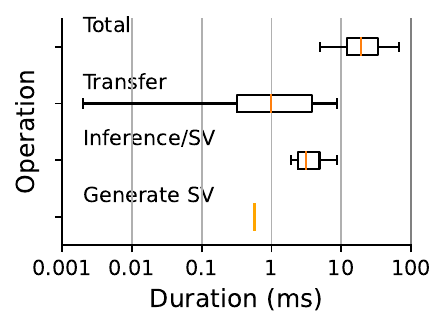}
     \vspace{-6.5mm}
    \caption{ML module related}
    \label{fig: ml_ops}
\end{subfigure}
 \vspace{-5mm}
\caption{Microbenchmarks and Overhead.}
 \vspace{-3mm}
\end{figure}

\subsection{Overall Performance} \label{eval: overall}
In this experiment, we enable both modules (e.g., reclaim pool and ML module) and evaluate the overall performance using a long trace.
We use the same server mentioned in Section~\ref{sec: motivation} with a Nvidia 3080 GPU. 
Specifically, we compare \ourSystem with LRU, LFU, FaasCache using a 6-hour traces distributed to 50 mobile users and 20 regular users (i.e., scenario 3).
The same set of traces are fed into four systems running on the same hardware testbed independently for evaluation. 
We set the maximum of containers to 32, and \ourSystem is configured with 24-8 (warm pool - reclaim pool) setup.
For other systems, we keep the default keep-alive period (i.e., 10 minutes). 
For \ourSystem, we set keep-alive policy as 5 minutes for both warm pool and reclaim pool.
During the evaluation, we also measure the CPU and memory utilization (the results are presented in the following subsection).

Figure~\ref{fig: 6 hrs exp} illustrates the number of cold start for each system. 
We calculate the number of cold start in every 20 minutes, and draw circles with the radius proportional to the number of cold start.
Basically, the larger circles are, the more cold starts occur in a 20-minute period.
All four systems have many cold starts in first 20 minutes because the warm pool is empty at the beginning.
However, \ourSystem has much better performance with much less cold start over time (i.e., the circle size is smaller than others).
Indeed, the number of cold start is much lower than other systems (e.g., 364 for FaasCamp versus 566 for FaasCache).

\subsection{Implementation Overhead} \label{eval: overhead}
\noindent\textbf{Micro-benchmarks.} The two modules of \ourSystem, reclaim pool and ML module, both introduces extra operations.
We employ a series of micro-benchmarks to measure the time consumption of multiple operations, including (1) cold-start container + make a checkpoint; (2) cold-start container without creating checkpoint; (3) move a container from the warm pool to reclaim pool (with the restore operation); (4) use a container from the reclaim pool (add namespace information); (5) use a container from the warm pool; (6) State vector creation for one container; (7) ML inference per state vector; and (8) network transmission for eviction results.

\begin{figure}[t]
\begin{subfigure}{.49\linewidth}
    \hspace{-3mm}
    \includegraphics[scale=0.59]{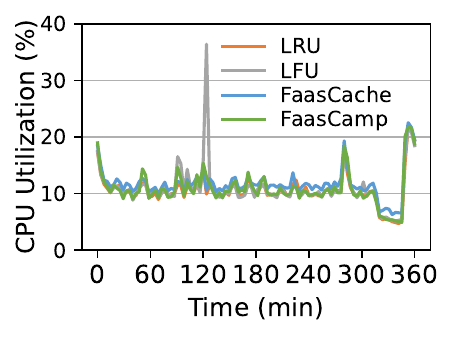}
     \vspace{-6.5mm}
    \caption{CPU}
    \label{fig: utilization}
\end{subfigure}
\hfill 
\hspace{-5mm}
\begin{subfigure}{.49\linewidth}
    \centering
    \includegraphics[scale=0.59]{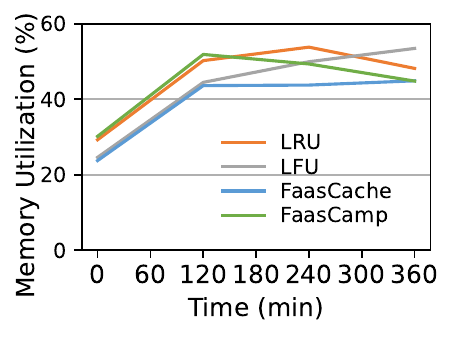}
    \vspace{-6.5mm}
    \caption{Memory}
    \label{fig: utilization-mem}
\end{subfigure}
 \vspace{-1mm}
\caption{System Utilization.}
\label{fig: util}
 \vspace{-4mm}
\end{figure}

\begin{table*}[t]
    \caption{The average warm rate, response time, and queue size of the different systems in the multi-invoker evaluations. High warm rate and low response time, queue size indicates high throughput.} 
     \vspace{-1.5mm}
    \centering
    \label{tbl: scalability}
    \begin{tabular}{p{14mm}*{9}{|p{10mm}}}
    \toprule
        \multirow{2}{*}{\textbf{System}} & 
            \multicolumn{3}{c|}{\textbf{Avg. Warm Rate}} & 
            \multicolumn{3}{c|}{\textbf{Avg. Response Time (s)}} & 
            \multicolumn{3}{c}{\textbf{Avg. Queue Size}} \\ \cmidrule(lr){2-4}\cmidrule(lr){5-7}\cmidrule(lr){8-10} &
            \hfil \textbf{S1} & 
            \hfil \textbf{S2} & 
            \hfil \textbf{S3} & 
            \hfil \textbf{S1} & 
            \hfil \textbf{S2} & 
            \hfil \textbf{S3} & 
            \hfil \textbf{S1} & 
            \hfil \textbf{S2} & 
            \hfil \textbf{S3} \\\midrule
        \hfil LRU & 
            \hfil 86.55 &
            \hfil 52.07 &
            \hfil 73.36 &
            \hfil 4.86 &
            \hfil 10.69 & 
            \hfil 10.4 &
            \hfil 4.87 &
            \hfil 5.73 & 
            \hfil 9.99 \\
        \hfil LFU & 
            \hfil 78.52 &
            \hfil 47.33 &
            \hfil 66.53 &
            \hfil 11.27 &
            \hfil 11.59 & 
            \hfil 16.02 &
            \hfil 10.32 &
            \hfil 6.27 & 
            \hfil 17.04 \\
        FaasCache & 
            \hfil 73.1 &
            \hfil 46.83 &
            \hfil 59.47 &
            \hfil 14.36 &
            \hfil 12.6 &
            \hfil 23.02 &
            \hfil 13.46 &
            \hfil 7.68 & 
            \hfil 25.7 \\
            \cellcolor[gray]{0.9} FaasCamp & 
            \cellcolor[gray]{0.9}     \hfil 94.85 &
            \cellcolor[gray]{0.9}     \hfil 70.32 &
            \cellcolor[gray]{0.9}     \hfil 91.13 &
            \cellcolor[gray]{0.9}     \hfil 0.59 &
            \cellcolor[gray]{0.9}     \hfil 7.97 & 
            \cellcolor[gray]{0.9}     \hfil 1.17 &
            \cellcolor[gray]{0.9}     \hfil 0 &
            \cellcolor[gray]{0.9}     \hfil 3.58 & 
            \cellcolor[gray]{0.9}     \hfil 0.22 \\
        \bottomrule
    \end{tabular}
 \vspace{-2mm}
\end{table*}

We measure the timing information using the model training workload for 200 times, as this is the most complicated workload from FunctionBench. Figure~\ref{fig: rp_ops} shows the duration of all operations involved with the reclaim pool, presented in milliseconds.
There is almost no difference between a warm start from the warm pool and a warm start from the reclaim pool. This is expected as the only overhead is adding namespace (tenant) information to containers, which is almost instantaneous.
Cold start a new image with a new checkpoint created consumes slightly more time than a normal cold start. 
Note that this is not the time for creating a checkpoint, which is done by a separate thread (thus has no impact on users).
As the overhead is only in millisecond level, we believe this overhead is acceptable.
Finally, the restore operation when moving a container from the warm pool to a reclaim pool definitely adds extra overhead (430 ms on average). However, this operation is transparent to users and thus will not affect user experiences. 

The timing measurements of machine learning related operations are presented in Figure~\ref{fig: ml_ops}.
All individual operations, (e.g., generating the state vector for each container, ML inference for each case, and the network transmission) take negligible overhead (less than 10 ms).
But the total time for making an eviction decision is decided by the number of available containers.
In our experiments, for a maximum of 8 containers, the average is only 38.63ms. 

\noindent\textbf{System Overhead.} We also compare the system overhead (e.g., CPU and memory consumption) of FaasCamp with other systems (e.g., vanilla Openwhisk and FaasCache). 
The CPU and memory utilization is collected through a system monitor \textit{sar} from the package \textit{sysstat}~\cite{sysstat}.
Figures~\ref{fig: utilization} and \ref{fig: utilization-mem} demonstrate the CPU load and the memory consumption for running the 6-hour trace. 
While the system utilization changes based on the number of concurrent requests, there is almost no difference for the CPU utilization among four systems. 
Also, FaasCamp shows a similar memory usage pattern as others, with the average memory consumption less than 1\% difference compared with LRU and LFU.
The results indicate that the overhead of FaasCamp is acceptable. 

\subsection{Multi-Invoker Performance} \label{eval: cluster}
Finally, we evaluate our system in a multi-invoker cluster setup, with three physical servers included. 
Each server runs one invoker on Ubuntu 20.04 with 16GB of memory. 
Among them, one is the main server hosting the controller and dispatching incoming requests to invokers based on the default load balancing strategy of OpenWhisk.
Similarly, we conduct 15 experiments with randomly selected traces for each scenario (as mentioned in \S~\ref{eval: method}) per system.
Each invoker is capped with 32 containers, and we choose 24-8 as the configuration for \ourSystem.

In realistic scenarios, flash events regularly occur, generating bursty incoming request patterns that are beyond the cluster's capacity. Requests are then queued, incurring queuing delays. To demonstrate that our system can achieve high throughput, we use three different metrics: (1) Warm rate; (2) Request response time; and (3) System queue size. 
Particularly, for each request, the response time includes three components: the initialization (cold start) overhead, the wait time in the queue (i.e., queuing delay), and the execution duration. 
The wait time represents the interval between when a request is received by the controller and when a container is provisioned by an invoker. 
A shorter response time indicates that the system is more efficient in provisioning containers, highlighting the effectiveness of the reclaim pool.

The system queue size (i.e., the number of queued requests) is another indicator of throughput: a smaller queue size suggests that the system can handle more requests within a given time frame.
Table~\ref{tbl: scalability} presents the results, demonstrating that \ourSystem outperforms other systems in all scenarios.


\section{Related Work}
\label{sec: RW}
\noindent\textbf{Cold Start Optimization.}
Extensive research has been conducted to optimize serverless computing~\cite{tariq2020sequoia,bhasi2021kraken,kannan2019grandslam,daw2020xanadu,stojkovic2023specfaas,roy2022icebreaker}, particularly the cold start process~\cite{singhvi2021atoll,Firecracker,akkus2018sand,ao2022faasnap,faascache,shillaker2020faasm}. 
SOCK~\cite{sock} and Catalyzer~\cite{Catalyzer} build template images to shorten the package download and install time.
Firecracker~\cite{Firecracker} and FaaSnap~\cite{ao2022faasnap} utilize lightweight VMs for improving function execution.
Our work can be combined with them to further improve the sandbox warm ratio.

Another common approach is to reduce the sandbox provisioning frequency~\cite{faascache,wild,li2022help}.
Romero et al.~\cite{romero2021faa} developed a transparent auto-scaling distributed cache: the cache is pre-warmed following data access patterns on Azure.
IceBreaker~\cite{roy2022icebreaker} dynamically warm up a function based on the function’s time-varying probability of the next invocation.
Unlike them, we adopt multiple optimizations from the cache to enhance serverless computing performance.

\noindent
\textbf{ML in Serverless Computing.}
ML has also be applied in optimizing serverless computing~\cite{9162876, schuler2021ai, yu2021harvesting, siren, 9749611, 8975850}.
For example, neural networks have been utilized to predict request so that a container can be pre-warmed to avoid cold start~\cite{9749611, 8975850}.
Others use reinforcement learning to adjust system parameters to maximum resource utilization~\cite{9162876, schuler2021ai, yu2021harvesting, siren}.
Unlike these prior efforts, \ourSystem develop a framework enabling ML-based container eviction, and any deep learning models can be integrated into our system.

\noindent\textbf{Approximating Bélády's Algorithm.} 
Many works attempt to approximate Bélády's algorithm using machine learning techniques in both the cache design and other computing scenarios. 
Both Hawkeye~\cite{hawkeye} and Glider~\cite{glider} construct neural networks using program counters as input features for cache replacement.
RLR~\cite{rlr} adopts reinforcement learning with neural network using time and frequency of cache lines.
Bélády's algorithm has also been utilized in guiding caching in the content distribution network (CDN)~\cite{yan2020rl,yan2022towards,relaxedBelady}.
Song et al.~\cite{relaxedBelady} developed Learning Relaxed Belady to mimic a Relaxed Belady algorithm in CDN cache design.
All these prior works indicate that approximating Bélády's algorithm is possible and feasible, and further motivate our works. 

\noindent
\textbf{Serverless Edge Computing.}
Serverless computing has been combined with edge computing in past years~\cite{aske2018supporting,baresi2019towards,baresi2019unified,xie2021serverless}.
Mittal et al.~\cite{mittal2021mu} adapted Knative~\cite{knative} to edge devices with limited computing resources, and Cicconetti et al.~\cite{cicconetti2020decentralized} proposed a decentralized architecture to manage edge computing nodes to mitigate network congestion.
Our work can be integrated with previous works to further improve the performance of serverless edge computing.


\section{Conclusion}
\label{sec: conclusion}
This paper presents \ourSystem, a serverless computing framework with a multi-tier warm pool design and ML-enabled invoker.
\ourSystem enables secure cross-tenant container sharing via the checkpoint and restore techniques.
It can improve the warm ratio compared with the existing singer-tier warm pool. 
Also, \ourSystem leverages deep learning techniques as the container keep-alive policy, guided by the optimal cache replacement algorithms. 
We have implemented \ourSystem and conducted extensive experiments on evaluating both performance and overhead using real-world traces and workloads. 
Our results show that the \ourSystem can enhance performance  with minimal overhead.


\section*{ACKNOWLEDGMENTS}
This work is partially supported by National Science Foundation (NSF) grants CNS-2054657 and OAC-2319975.

\bibliographystyle{IEEEtran}
\bibliography{ref.bib}

\end{document}